\documentclass[a4paper]{jpconf}
\usepackage{graphicx}

\usepackage{citesort}
\begin{document}

\title{Galactic Propagation of Cosmic Rays from Individual Supernova Remnants}

\author{Nils Nierstenhoefer, Philipp Graeser, Florian Schuppan and Julia Becker Tjus}

\address{Ruhr Universit\"at Bochum, Universit\"atsstra\ss e 150, D-44780 Bochum}

\ead{nils@tp4.rub.de}

\begin{abstract}
It is widely believed that supernova remnants are the best candidate sources for the observed cosmic ray flux up to the knee, i.e.\ up to $\sim$PeV energies. Indeed, the gamma-ray spectra of some supernova remnants can be well explained by assuming the decay of neutral pions which are created in hadronic interactions. Therefore, fitting the corresponding gamma spectra allows us to derive the spectra of cosmic rays at the source which are locally injected into our Galaxy. Using these spectra as a starting point, we propagate the cosmic rays through the Galaxy using the publicly available GALPROP code. Here, we will present first results on the contribution of those SNRs to the total cosmic ray flux and discuss implications.
\end{abstract}

\section{Introduction}
Many popular theories for the origin of \textit{Galactic} cosmic rays (CRs) up to the knee-region ($\sim$PeV) are based on the idea of shock acceleration of charged particles in supernova remnants (SNRs)~\cite{FermiShockAcc49,BlandfordOstriker1978,Hillas2005Rev}. That SNRs indeed provide the required energy budget to be the birthplace of the Galactic cosmic ray population has firstly been pointed out in~\cite{ZwickyBaade1934}. 
\newline
Recent indirect observational techniques as used by the instruments H.E.S.S., MAGIC, VERITAS and MILAGRO allow for the observations of very high energy (VHE) gamma-rays from SNRs up to $\sim$100~TeV. Including radio and X-ray observations, it is possible to test and distinguish models of leptonic or hadronic origin of those gamma-rays as has been done systematically for 24 well-identified SNRs in~\cite{mandelartz_statistical_2013}. In particular, this kind of analysis provides a parametrization $j_{p}(T)$ of a potential population of CR protons inside the SNRs. This indirect approach to derive $j_{p}(T)$ is useful as the CRs trajectories, masses and energies are altered during their Galactic propagation - thus crucial \textit{direct} information is lost. This is different for the VHE gamma-rays which propagate rectilinear with no significant energy losses through the Galaxy.
\newline
For the same reason, the effects of Galactic CR propagation have to be taken into account if one aims on testing whether the source spectra $j_{p}(T)$ of SNRs can explain the CR flux as measured on Earth. To do so, an altered version of the publicly available GALPROP framework~\cite{strong_new_1998} for Galactic propagation is used, in which CR nuclei are injected according to the source spectra derived in~\cite{mandelartz_statistical_2013}. It needs to be taken into account that these spectra represent currently observed SNRs, and thus only a short temporal phase of cosmic ray injection: while a single SNR is only active for about $10^{4}$~years, cosmic rays diffuse in the Galaxy for about $10^{6}-10^{7}$~years. In order to have a time-averaged sample of SNRs, we assume that the CR spectra as derived from the currently observed SNRs are a representative subsample of SNR spectra and luminosities. We then perform multiple simulations with SNR locations randomly drawn from the spatial distribution of \cite{case_new_1998}, which represents a model for the distribution of SNRs in the Galaxies, with an enhanced density near the Galactic Center.
\newline
The main goal of this paper is to describe our current simulation approach and to present an exemplary simulation which is focusing on the investigation of the shape of the predicted CR proton flux at Earth. In particular, we emphasize the fact that we do \textit{not} perform any optimization of model parameters here and that we present the method of our approach with only preliminary physics results in these proceedings. All details of the physics questions, including a parameter study will be discussed elsewhere in the future. One of the primary, final goals of developing this method is the evaluation of the overall flux normalization.

Finally, in these proceedings, we present improvements of the simulation approach and their impact on the interpretation of the results are outlined.  
\newline

\section{Source Spectra\label{sct:SourceSpec}}
The modeling of the electromagnetic spectrum of 24 SNRs has been performed in ~\cite{mandelartz_statistical_2013}.  Out of those,  a significant contribution of gamma-rays from hadronic interactions with a subsequent $\pi_{0}$-decay are found for 21 SNRs (see table 3 in~\cite{mandelartz_statistical_2013}). These 21 source spectra $j_{p}(T)$ are the input for the analysis which is presented in what follows. 
\newline
More precisely, the authors of ~\cite{mandelartz_statistical_2013} apply a one zone model for the SNR and provide a parametrization of a simple power law for the proton spectra in momentum $p=\sqrt{mc^2-(T+mc^2)^2}$, which translates into an energy spectrum of
\begin{equation}
j_{\rm p}(T)=a_{p}\ \sqrt{\frac{T^2+2Tmc^2}{T_{0}^2+2T_{0}mc^2}}\ \frac{T+2Tmc^2}{\sqrt{T^2+2Tmc^2}}\ \tanh\left( \frac{T}{T_{\min}} \right)\ \exp\left(-\frac{T}{T_{\max}} \right).
\label{eq:SourceSpec_Mandelartz}
\end{equation}
Here, $T, a$ and $m$ denote the kinetic energy, the normalization constant and the proton mass, respectively. A smooth low energy cut off at $T_{\min}=10$~MeV and its high  energy counterpart at $T_{\max}$ is included through the factors four and five in equation (\ref{eq:SourceSpec_Mandelartz}). In this paper, the focus is on the CR energy range from $\sim$GeV to $\sim$PeV. Hence, the low energy cut off is not applied.
\newline
It should be noted that for example 274 SNRs are listed in the well known Greens catalogue ~\cite{green_revised_2009} while only $\sim$20 SNRs have been detected in gamma-rays. Correspondingly, the prediction of a too low CR-flux is expected here. 
\newline
Only SNRs that are observed in MeV-GeV or GeV-TeV are used in the analysis of ~\cite{mandelartz_statistical_2013} and are included in our analysis. Therefore, the way of selecting sources for the observation with e.g.\ Cherenkov telescopes, may induce a selection bias (e.g.\ favouring flat, luminous SNRs). Therefore, the set of 21 SNRs from ~\cite{mandelartz_statistical_2013} is not statistically complete and, consequently, the presented results need to be interpreted with adequate care. Future data taken with the instruments CTA and HAWC will provide a more complete picture of gamma-ray emitting SNRs in the Galaxy and this study can be updated accordingly. 

\section{Propagation \label{sec:Propagation}}
The effects of Galactic propagation on the source spectra $j_{p}(T)$ was simulated using version 54 of the open source C++-framework GALPROP~\cite{strong_new_1998,Galprop_Web_Standford,strong_galprop_2011}. The code was extended to allow for particle injection according to  $j_{p}(T)$ as parametrized in equation (\ref{eq:SourceSpec_Mandelartz}). Additionally, a procedure was implemented that allows to randomly select a position for a set of SNRs from a predefined distribution of matter within the Galaxy. 
\subsection{Normalization}
First of all it should be noted that the CR normalization in \textit{standard} GALPROP applications is usually fixed by globally scaling the calculated cosmic ray density such that the observed CR flux at Earth is met. In the approach presented in these proceedings, the idea is to fix the normalization by the rate of particle injection of the individual SNRs into the Galaxy\footnote{In the exemplary simulation discussed later in this paper the predicted CR flux is normalized to experimental data. The more extensive subject of modeling the CR normalization as a direct consequence of the source spectra $j_{p}(T)$ will be discussed in a future paper.}. Hence, from the technical viewpoint a key ingredient was to rewrite the source spectrum  $j_{\rm p}(T)$ in terms of the \textit{internal} GALPROP units. A pragmatic approach to determine the needed conversion factor is to compare the calculation of the luminosity $L$ in GALPROP with the corresponding integral expression on the basis of $j_{p}(T)$
\begin{equation}
L = \alpha R^2_{\rm SNR}c \int_{10~\rm{MeV}}^{1e9~\rm{MeV}}\ dT\ \beta\ T\ j_{T}(T). 
\label{eq:LuminosityGeneral1}
\end{equation}   
Here, $R_{\rm SNR}$ is the radius of the SNRs. Note that $\alpha =1$ corresponds to the case where the CR particles stream out of the SNR in radial direction. In contrast to that $\alpha =$1/2 would arises from an averaging assuming that the CRs move in random directions inside the SNRs. The aforementioned comparison leads to the following relation between the initial source function $q_1(p(T))$ as implemented in GALPROP\footnote{Further details can be found in the GALPROP documentation~\cite{strong_galprop_2011,Galprop_Web_Standford} or by directly checking the source code file cr\_luminosity.cc.} 
and the spectrum parametrization  $j_{\rm p}(T)$ 
\begin{equation}
q_1(p(T))= \alpha \frac{c^{2} R_{\rm SNR}^{2} \beta^{2}}{4\pi V_{\rm grid}}\ j_{\rm p}(T).
\label{eq:ConversionISF}
\end{equation}
Alternatively, the luminosity can be derived from the total energy $E_{\rm tot}$ of protons in the SNR
via $L=E_{\rm tot}/\tau$ with a time-scale $\tau$, representing the distribution of the total energy over the total  lifetime of the remnant,
\begin{equation}
L =  \frac{1}{\tau} \frac{4\pi}{3}R^3_{\rm SNR} \int_{10~\rm{MeV}}^{1e9~\rm{MeV}}\ dT\ T\ j_{T}(T). 
\label{eq:LuminosityGeneral2}
\end{equation}
Using this expression for the luminosity, one finds an alternative expressions\footnote{Equation (\ref{eq:ConversionISF}) has been applied in this paper.} for equation (\ref{eq:ConversionISF})
\begin{equation}
q'_1(p(T))= \frac{\beta c R_{\rm SNR}^{3}}{3 V_{\rm grid} \tau}\ j_{\rm p}(T).
\label{eq:ConversionISF_alter}
\end{equation}
Assuming that $E_{\rm tot}$ is the energy of the SNR converted into protons and $\tau$ is the life time of the SNR, $L=E_{\rm tot}/\tau$ can be interpreted as the average luminosity in CRs.

It is a typical approach in gamma-ray astronomy to derive the total cosmic ray energy budget in order to estimate the SNRs possible contribution to the total cosmic ray budget. A back-of the envelope calculation predicts that the observed cosmic ray luminosity of the Galaxy (about $3\cdot 10^{40}$~erg/s) can be reproduced if on average, $10^{50}$~erg are going into a single SNR at a Supernova rate of $1/(100$~years$)$. Here, it is assumed that cosmic rays are injected into the interaction region continuously at a constant rate during the lifetime $\tau$ with the total energy going into cosmic rays conserved over time. It is clear that this is a simplifying assumption, as it is known that at least the energy spectrum is changing with time, in particular concerning the reduction of maximum energy, see e.g.\ \cite{cox1972}. For energy spectra steeper than $E^{-2}$, the total energy budget is dominated by the lower integration threshold, so effects from this temporal development should be relatively small. It is also specified in \cite{cox1972} that the total energy of the SNR is decreasing with time due to cooling effects. This would mean that, if we assume a constant fraction of the SNR energy going into cosmic rays at a given time, the actual average luminosity of cosmic rays would be underestimated in particular for old remnants. It is not clear, however, if the fraction of energy going into cosmic rays is actually constant over time or if it actually decreases with the available energy budget. Thus, we judge that in first order approximation, it seems reasonable to estimate the total energy of the remnant from the given value, keeping in mind the above discussed uncertainties.
\newline
In these proceedings, we refrain from discussing the {\it global} normalization and normalize the final spectrum to the observed flux. The normalizations of the individual SNRs relative to each other are fixed using equation \ref{eq:ConversionISF}.
\subsection{Including CR Nuclei}
Although the source spectra taken from~\cite{mandelartz_statistical_2013} are only provided for CR protons, it is possible to include CR nuclei in GALPROP simulations (see chapter 5.5 in ~\cite{strong_galprop_2011,Galprop_Web_Standford}). To do so, the initial source function of nuclei $q_{A}(p_{A})$ with mass number $A$ and momentum $p_{A}$ is related to $q_{1}(p_{1})$ by the relative abundance $X$ according to
\begin{equation}
X=\frac{Aq_{A}(p_{A})}{q_{1}(p_{1})}.
\label{eq:RelAbundanceAndNuclei}
\end{equation}
In this context two remarks have to be made: 
\begin{enumerate}
\item Due to the high energy cut off in equation (\ref{eq:SourceSpec_Mandelartz}) $X$ is not independent of the energy in what follows. 
\item Including CR nuclei injection, the total energy in hadrons of the SNRs is artificially increased. Note, one can approximately correct the energy budget derived in ~\cite{mandelartz_statistical_2013} by down-scaling the proton normalization $a_{p}$ in equation (\ref{eq:SourceSpec_Mandelartz}) appropriately.
\end{enumerate}
Here, simulations are performed for all nuclei, but the resulting energy spectra are only discussed for protons. In the future, the  heavy nuclei spectra will be discussed as well in order to investigate other questions connected to cosmic ray observations, sources and transport.

\subsection{Random Positions, Mapping to Closest Gridpoint and Averaging \label{SubSec:Propagation:RandPosClosGriPAverage}}
The idea behind GALPROP is to numerically solve the transport equations that governs the propagation of CRs in the Galaxy ~\cite{strong_new_1998}. This requires the definition of a spatial simulation grid on which the particle densities are calculated. Here, we select random position for the SNRs and map them to the closest grid point. As the grid size is larger than the extensions of the SNRs, this implies that currently the SNRs are treated as point like CR sources.
\newline
We finally calculate and present average values - e.g.\ the average proton flux $<dF/dT>$ or the Boron to Carbon ratio $<B/C>$ - from $N$ sets of 21 SNRs. It is  this averaging which allows one to derive a realistic prediction for CR observables.
\newline
Note that the variance of CR observables e.g.\ Var($dF/dT$) or related quantities cannot be conclusively calculated within our current simulation procedure. The reasons for that and and a method to overcome this issue in the future is outlined in section \ref{sct:discussion}.
\subsection{GALPROP Settings}
As a starting point we use the \textit{standard} settings as suggested by the GALPROP web-run application web page~\cite{strong_galprop_2011,Galprop_Web_Standford}. One change is applied to the standard settings: The size of the galaxy has been reduced in our example simulation to -10~kpc$<x,y<$10~kpc - to allow faster simulations. In particular we do not aim for optimizing free model parameters to improve the agreement with experimental data at this point. Here, we mainly intend to outline our computational approach and to show exemplary results - further results (including more detailed parameters studies)  will be reported elsewhere in the future. 

\section{Exemplary Simulation}
In this section an exemplary simulation is presented. The basic parameters for the underlying GALPROP runs are mainly the ones suggested by the GALPROP web-run web page~\cite{strong_galprop_2011,Galprop_Web_Standford} (cf. section \ref{sec:Propagation}). In this example $N$=100 sets of 21 SNRs with random positions according to the SNR-distribution of ~\cite{case_new_1998} have been simulated. In what follows, the  bracket-notation ($<...>$) indicates that the predicted observables that can be compared to the observations are calculated as an average of $N$=100 simulated sets of SNRs with random positions (cf. section \ref{sec:Propagation} for details). 
\begin{figure}
  \begin{center}
    \includegraphics[width=1.0\textwidth]{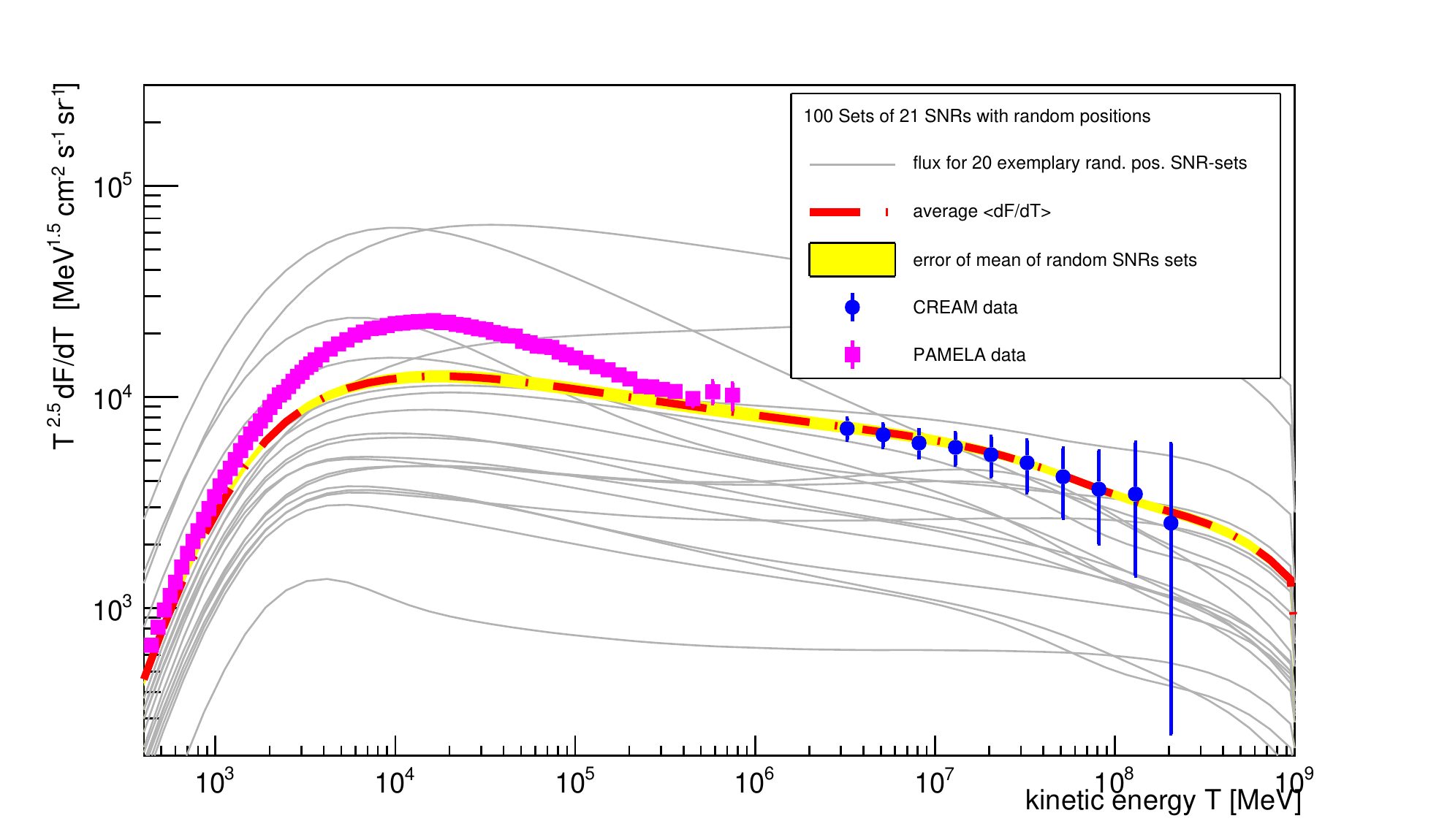}
  \end{center}
  \caption{\label{fig:GalComp:Spec:NormFromEtot}20 out of 100 proton spectra $dF/dT$ of random-position sets of 21 SNRs are exemplarily presented in grey. Mean values $<dF/dT>$ (red) and their errors (yellow band) are displayed. Blue dots show measured data by CREAM ~\cite{yoon_cosmic-ray_2011}, the magenta curve presents the observations by PAMELA ~\cite{adriani_pamela_2011}. Note that that $<dF/dT>$ is normalized to fit the CREAM data point at lowest energy.}
\end{figure}
\newline
The predicted flux $<dF/dT>$ of cosmic ray protons is compared to experimental data from CREAM ~\cite{yoon_cosmic-ray_2011} and PAMELA~\cite{adriani_pamela_2011} in Fig.~\ref{fig:GalComp:Spec:NormFromEtot}. Here, the statistical and systematic experimental uncertainties are displayed. Additionally, $<dF/dT>$ has been normalized to CREAM data at 3.25~TeV to allow for a comparison of the spectral behaviour. Note that the proper solar modulation potential of 550~MV linked with the PAMELA measurements~\cite{adriani_pamela_2011} is included in Fig.~\ref{fig:GalComp:Spec:NormFromEtot} according to~\cite{1968ApJ...154.1011G}.  
\\
We find that the spectral behavior of the measured proton spectrum is well described in the energy range of the CREAM experiment, but a discrepancy with the PAMELA data which is most obvious at 10~GeV is apparent. As we derive the individual proton spectra from gamma-ray data, we expext that best agreement should be present in the statistically well-measured range, which is around $1-100$~GeV for Fermi and $100$~GeV-$10$~TeV for IACTs. Assuming that about $10\%$ of the cosmic ray energy is going into photons, this would mean a good description between $10$~GeV and $100$~TeV. The disagreement is therefore not arising from any extrapolation effect.
\newline   
\begin{figure}
  \begin{center}
    \includegraphics[width=1.0\textwidth]{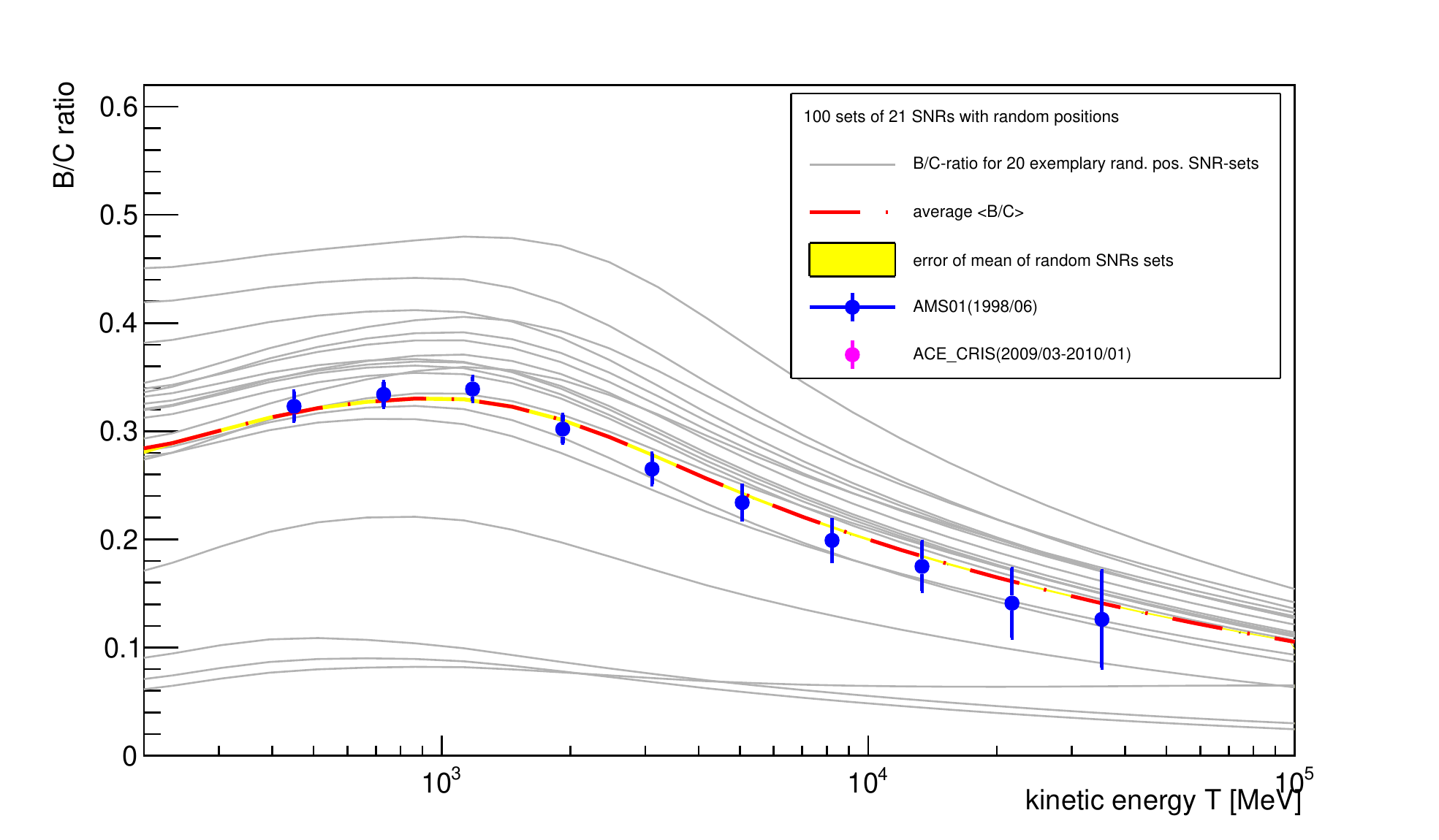}
  \end{center}
  \caption{\label{fig:GalComp_BCRatio:NormFromEtot}20 out of 100 $B/C$-ratios of random-position sets of 21 SNRs are exemplarily presented in grey. Mean values $<B/C>$ (red) and their errors (yellow band) are displayed. Blue dots show measured data from AMS-01~\cite{aguilar_relative_2010}.}
\end{figure}

As a cross-check, we also include the simulation of heavy nuclei and find that the simulated $<B/C>$-ratio is in good agreement with data measured by AMS-01 as can be seen in Fig.~\ref{fig:GalComp_BCRatio:NormFromEtot}.  

\section{Discussion\label{sct:discussion}}
In general the good agreement between $<B/C>$ and experimental data indicates that the used simulation setup is well suited to model Galactic CR propagation. Taking into account that the observed proton spectrum is well reproduced for energies larger than $\sim$1~TeV, supports the famous theory that SNRs are the birthplace of the Galactic CRs up to $\sim$PeV energies.

These first results show that there is a \textit{possible} discrepancy between $<dF/dT>$ and the observations by PAMELA which is most apparent at kinetic energies of $T\sim10~$GeV, using GALPROP standard parameters. In the following, we will discuss both statistical and physics reasons for the disagreement.

\textit{Statistical reasons} may contribute to the disagreement: our current simulation approach predicts the average CR observable, e.g.\ $<dF/dT>$, but not the proper corresponding variance, e.g.\ $Var(dF/dT)$ (see the discussion later on). However, the variance or alternative statistical measures are needed to fully quantify whether the aforementioned discrepancy of $<dF/dT>$ is statistically significant.
To quantify the agreement between the predicted CR observables statistically, a measure of the \textit{variance} is needed. The variance of the CR observables could be calculated and presented here along with the mean value. However, it should be noted that this variance has a limited statistical interpretation in the current simulation approach. It merely measures the spread of the simulated CR observable which is induced by randomly selecting positions for the 21 SNR in the Galaxy. In particular, this variance would decrease if the number of SNRs would be increased. As more than 21 SNRs are expected to contribute to the CR flux the variance calculate here would presumably be an upper limit for the true variance. 
In addition, the calculation of the true variance would require to include temporal aspects of the SNRs - such as their production rate and lifetime - into our simulation approach. This could be done as a generalization of our procedure to include spatial aspects via the selection random positions and mapping of SNRs to the closest grid point on the spatial simulation grid (cf. subsection \ref{SubSec:Propagation:RandPosClosGriPAverage}). As similar issues are already addressed in the GALPROP manual ~\cite{strong_galprop_2011,Galprop_Web_Standford} (see chapter 6), an implementation including the temporal aspects of SNRs should be feasible.
A nearby source contributing to the measured CR flux is another possible explanation for the disagreement in the proton spectra. Extending the method to provide a meaningful measure of the variance induced by spatial and temporal aspects of the SNRs will help to estimated the theoretical probability that such a constellation is realized in nature.

Possible \textit{physics} reasons which can contribute to this potential disagreement are the following:
\begin{enumerate}
\item  By using the standard GALPROP settings, we use a diffusion scalar with Kolmogorov-like diffusion, leading to a general steepening of the energy spectrum of a distant source by a power of $\sim 1/3$. Measuerments like the one of the Boron-to-Carbon ratio show that the scalar approximation would even allow for a larger steepening \cite{obermeier2012}. This would presumably give a better fit at low energies, while a discrepancy would show towards higher energies instead.
\item The latter behavior could be explained by missing, flat SNRs, which are not seen yet due to missing sensitivity at the highest energies: The SNR sample extracted from~\cite{mandelartz_statistical_2013} is not statistically complete (compare section \ref{sct:SourceSpec}).  It is expected to have a more complete sample of SNRs in the future, with available CTA and HAWC data. Also, future results from IceCube, KM3NeT and IceCubeGen2 will presumably help to clarify which part of the gamma-ray signal actually is of hadronic nature. Correlation studies between gamma-rays and ionization signatures can also contribute to identifying those sources that are hadronically dominated, see e.g.\ \cite{becker2011,schuppan2012,schuppan2014}.
\item Another possibility would be a diffusion scalar with a power law behavior broken at around $100$~GeV, as already discussed in \cite{Galprop_Web_Standford}.
  \item Finally, the full treatment of the diffusion process with a tensor might change the picutre. Here, the difficulty is that the structure of the tensor itself is not very well known. First investigations of tensor diffusion in the Galaxy are presented inplemented in the DRAGON code (\cite{dragon}, so far optimized to the propagation of electrons \cite{gaggero2013}) and the PICARD code (\cite{kissmann2014}, published with first test examples).
\end{enumerate}

We already stressed that the parameters which define the framework for the Galactic propagation have not been optimized in our example application. Especially, using a broken or steeper power law for the diffusion coefficient as function of energy may provide a handle to reproduce the shape of the experimental proton spectrum at $T\sim$GeV.  

\section{Conclusion}
Based on the publicly available GALPROP code for Galactic CR propagation, a procedure is introduced to predict CR observables for a given set of hadronic source spectra from SNRs. This method is based on calculating the average over multiple GALPROP run using sets of SNR with positions randomly distributed according to a given spatial distribution.
In these proceedings, we illustrate the current procedure in which we use the proton spectra of 21 SNRs inferred from the corresponding gamma-ray measurements as presented in ~\cite{mandelartz_statistical_2013}. While the $<B/C>$ ratio is well reproduced in this simulation we find a discrepancy for the predicted proton flux $<dF/dT>$. As we only present an exemplary study with no parameter study at this point, we do not draw any conclusions from these findings yet. The aim of these proceedings is the presentation of the method itself. In the future, we will present a full parameter study, including the discussion of the normalization of the spectrum as well. The latter aspect will contribute to answering the question if those SNRs observed in gamma-rays are representative to explain the cosmic ray energy budget as observed at Earth.

In the future, we particularily plan to extend our procedure to (I) include temporal aspects of SNR evolution, (II) include additional SNRs and (III) implemented on an inhomogeneous simulation grid.

On longer terms we intend to perform Monte-Carlo (MC) simulations of the propagation of Galactic CRs. As a basis, we suggest to use the publicly available CRPropa MC-framework to study the propagation of ultra-high energy cosmic rays in extra galactic environments~\cite{2013APh....42...41K}. Especially the redesigned object orientated structure of the upcoming version 3.0 of CRPropa seems to allow for an easy extensions for the propagation of galactic CRs ~\cite{2013arXiv1307.2643A}. With today's computer technologies, a Monte Carlo treatment of Galactic CRs down to T$\sim$10-100~TeV seems possible with reasonable run times. For lower energies it may be sufficient to switch to a diffusive approximation to avoid the time intensive numerical solution of the equation of motion in the galactic magnetic field.

\section*{Acknowledgments}
This work was supported by the DFG research unit FOR1048 on {\it Instabilities, Turbulence and Transport in Cosmic Magnetic Fields}, part B4 ({\it Transport of cosmic rays from supernova remnants through the Galactic Magnetic Field}) and by the research department on plasmas with complex interactions (Bochum). NN would like to thank A.W.\ Strong and C.\ Evoli for insights about the GALPROP code. Furthermore, we acknowledge helpful discussions with B.\ Eichmann.  

\section*{References}
\bibliography{refs_julia}

\end{document}